\documentstyle[prb,aps,multicol,psfig]{revtex}
\begin{document}
\draft
\tighten

\title{Erratum: Kinetic Theory of Spin Coherence of Electrons in Semiconductors 
[Journal of Superconductivity {\bf14},
245 (2001)] }
\author{M. W. Wu}
\address{Department of Applied Physics,  University of Tokyo,  7-3-1, Hongo Bunkyo-ku, 
Tokyo 113-8656, Japan}
\maketitle

\bigskip

There are two typos in my recent published paper:\cite{wu2} (i) The second term on the right hand side 
of Eq. \ (8) should appear as
\[
+\sqrt{2}\sum_{\stackrel{k k^\prime q}{\sigma}}U_{k^\prime,k^\prime-q}c^\dagger_{c k+q\sigma}c^\dagger_{vk^\prime-q\sigma}c_{vk^\prime-\sigma}c_{ck-\sigma}\ ,
\]
which describes an electron in the  CB with spin $-\sigma$ and momentum $k$ being scattered to
the CB with opposite spin $\sigma$ and momentum $k+q$, and at the same time,
an electron from the VB with spin  $-\sigma$ and 
momentum $k^\prime$ being scattered  to the 
band with  momentum $k^\prime-q$ and spin $\sigma$.
 (ii) The left hand side of Eq.\ (17) should be $\frac {1}{2\tau_s^\prime(k)}$ instead of $\frac {1}{\tau_s^\prime(k)}$.

Both are just typos and the calculations in the paper are based on the correct 
formulism.

%\end{multicols}
\end{document}